\documentclass[
    ,final            
  ]
  {aipproc}
\layoutstyle{8x11single}

\begin{document}
\newcommand{\pzero}{$\pi^{\rm{0}}$}

\title{SciBooNE's neutral current \pzero  production measurements}


\author{Yoshinori Kurimoto for the SciBooNE collaboration}{
  address={Kyoto University, Sakyo-ku, Kyoto, Japan}
}



\begin{abstract}
 The SciBooNE Collaboration has measured neutral current neutral pion production by 
the muon neutrino beam at a polystyrene target ($\rm C_{8}H_{8}$).  
We obtained  $(7.7 \pm 0.5({\rm stat.})^{+0.4}_{-0.5}({\rm sys.})) \times 10^{-2}$
as cross section ratio of the neutral current neutral pion production to total charged current cross section
at the mean neutrino energy of 1.16 GeV. This result is consistent with the Monte Carlo prediction based on 
the Rein-Sehgal model 
\end{abstract}

\maketitle

\section{Introduction}
In this paper, neutral current neutral pion production by muon 
neutrinos (NC\pzero) is defined as  a neutral current
interaction by muon neutrinos where at least one \pzero
is emitted in the final state from the target nucleus.

NC\pzero is a potential major background to the $\nu_{e}$ 
appearance search, which is the primary purpose of  
modern neutrino oscillation experiments such as the T2K
experiment \cite{Itow:2002rk}. This is because the gamma ray
from a \pzero mimic an electron from $\nu_{e}$ interaction 
in a detector such as SuperKamiokande used for the T2K experiment
as in many NC\pzero events both gamma rays are not resolved 
and give a single electron-like ring.
For this reason, a precise measurement of the NC\pzero
cross section is essential. For the T2K experiment, a 10 \% 
uncertainty on this cross section is desired. 

\section{Experimental setup}
The SciBooNE experiment \cite{Hiraide:2009} uses 
the Booster Neutrino Beam (BNB) at Fermilab. The primary proton
kinetic energy is 8 GeV and the neutrino flux is dominated by muon 
neutrinos (93 \% total). The flux-averaged mean neutrino energy
is 0.7 GeV.
The SciBooNE detector is located 100 m downstream from the neutrino
production target. The detector complex consists of three sub-detectors;
a fully active fine grained scintillator tracking detector ($SciBar$) 
\cite{Nitta:2004nt}, an electromagnetic calorimeter ($EC$)
 \cite{Buontempo:1994yp}
and a muon range detector ($MRD$). The SciBar detector
consists of 14336 extruded plastic scintillator strips. 
The scintillators are arranged  vertically and horizontally
to construct a $3 \times 3 \times 1.7 {\rm m}^{3}$
 volume with a total mass of 15 tons
The EC is installed downstream of SciBar
 to measure \pzero and the intrinsic ${\nu}_{e}$
contaminlations. The EC is a ``spaghetti'' type
calorimeter made of 262 modules comprised of 1~mm diameter
 scintillating fibers embedded in lead foil. The modules construct one vertical
and one horizontal plane, and each plane has 32 modules.  The EC
has a thickness of 11 radiation lengths along the beam direction. 
The MRD is located downstream of EC in order to measure the momentum of muons up to 
1.2 GeV/c with range. The experiment took both neutrino and antineutrino data from June
2007 until August 2008. In total $2.64 \times 10^{20}$POT (protons on target) were 
delivered to the berylium target during the SciBooNE data run. After beam and detector 
quality cuts, $2.58 \times 10^{20}$POT are usable for physicas analysis;
$0.99 \times 10^{20}$POT for neutrino data and $1.53 \times 10^{20}$POT for antineutrino data.
Preliminary results from the full neutrino data sample are presented in this paper.

\section{Analysis}
\subsection{Event Reconstruction}
The first step of the event reconstruction is to search for
two-dimensional tracks in each view of SciBar using a cellular
automaton algorithm \cite{Maesaka:2005aj}.  
Three dimensional tracks are reconstructed by matching
the timing and edges of the two dimensional tracks. 
three-dimensional reconstructed track $3D~track$ here after.

In order to improve the reconstruction of gamma rays, we 
introduced $extended~track$. Extended tracks are reconstructed based on
3D tracks. There are two steps to reconstruct extended tracks. The first step
is merging two 3D tracks on a common straight line. Because some part of
single gamma rays are broken into two clusters in SciBar and result
in two 3D tracks. Such two 3D tracks are handled as single extended track 
after merging.
The second step is collecting hits around merged 3D tracks. 
This is because electromagnetic showers sometimes make hits far from the main part
of showers and these hits are not associated to 3D tracks. 

Some of gamma rays observed in SciBar have energy deposit in EC due to
leakage. After event reconstruction in SciBar we search for $EC~clusters$
 (the collection of continuous hits in EC) pointed by 3D tracks in SciBar. 
We call such a EC cluster as a $matched~EC~cluster$. 
\subsection{Event Selection}
\label{subsec:evsec}
The clearest feature of the NC \pzero production is
two gamma rays from {\pzero}s. While the main background
events are divided into two categories; the $internal$ background
and the $external$ background. In the $internal$ background events,  
the neutrino interactions in SciBar produce secondary particles
but the interaction modes of them are different from the NC\pzero
interaction. The interaction mode in the $internal$ background 
are mainly charged current interaction. The $external$ backgrond
 is particles coming from the
outside of the detectors. There are two type of the external 
background;accidental cosmic rays and $dirt$ events. The contribution 
of accidental cosmic rays in any event samples is small and 
estimated  by data taken during off-beam timing. Hence, our data 
shown here is after subtraction of the contribution of accidental cosmic
ray. In $dirt~events$, neutrinos interact
with materials such as wall of experimental hall and produce
secondary particle which make hits in SciBar. The event
selections for NC neutral pion production were developped
for selecting two gamma rays but rejecting these backgrounds.

\subsubsection{Pre-selection}
\label{sssec:pre}
We use events with more than one 3D tracks to choose two gamma rays events.
In addition, we reject events if there are hits at the first layer of SciBar
and the timing difference between these hits and the 3D tracks is less than
100 nsec. This reject dirt events where charged particles from outside 
of the detectors come to SciBar.
\subsubsection{Rejection of the side escaping 3D tracks}
\label{sssec:noside}
We reject events with 3D tracks escaping from the side of SciBar.
Because most of such 3D tracks are muons produced in the charged current
events. After this selection there are still muons stopping in SciBar or
escaping from the donwstream of SciBar. These muons are rejected 
in other selections described later.
\subsubsection{Decay electron rejection}
\label{sssec:decay_electron_rejection}
 To reject muons stopping in SciBar, we use the electrons from muon decay.
 Since most of the decay electrons are not reconstructed as 3D tracks 
 due to their low energy, we search the delayed hits at the edges of 3D tracks.
 We search the maximum timing difference between the initail edge and end 
 edge of 3D tracks.
 If a muon decays to electron in a event, the maximum timing difference is
 corresponding to the muon life time (${\tau}_{\mu}~=~2.2~{\mu}~{\rm sec}$).
 Since most
 of events without decay electrons have the maximum timing difference less than 100 ns,
 events with the maximum timing difference less than 100 ns are selected.  
\subsubsection{Track disconnection cut}
 Charged current events often have multiple 3D tracks with a common vertex
 while two gamma rays from {\pzero}s usually are isolated from each other. 
 Hence, the distance between two tracks is used to separate two gamma rays from
 charged current event. We search the minimum distance between the edges of all
 3D tracks. If there are two particles with a common vertex, the minimum distance 
 is close to zero. 
 Events with the minimum distance greater than 6 cm are selected.
\subsubsection{Proton rejection}
 Since protons give a large energy deposit in SciBar, 
 the proton track is identified from other particles. 
 Using this information, we require events to have 
 at least two 3D tracks both of which are not protons.
 By this requirement, we reject charged current 
 events furthermore (for example, events with muon and protons
 at the final state). We define $Muon~confidence~level$ (MuCL)
 by using $dE/dx$ information as shown in \cite{Hiraide:2009}
 the MuCL is close to the maximum value (1) for muons
 and the minimum value (0) for protons.
 We define a track with MuCL greater than 0.03 as a
 $non{\rm -}proton{\rm -}like$ track. Events with at
 least two non-proton-like tracks are selected.
\subsubsection{Electron Catcher Cut}
 \label{sssec:eccut}
Matched EC Clusters are used to reject muons escaping
from the donwstream part of SciBar.
The two values are used. The one is the energy deposit in
 the upstream (vertical) EC cluster called $Edep_{\rm upstream}$
 and the other is the energy ratio of the downstream (horizontal)
 EC cluster to the upstream EC cluster called ${\rm R}_{\rm energy}$. 
If there are no matched EC clusters, $Edep_{\rm upstream}$ is set
 to zero and ${\rm R}_{\rm energy}$ are left undefined.
Since muons tend to penetrate material than $\gamma$s, the energy deposit
of muon at both upstream and downstream plane are close to each other 
($Edep_{\rm upstream} \sim 50~{\rm MeV}$, ${\rm R}_{\rm energy} \sim 1$). 
While gamma rays stop in  the short range after conversion with large
 energy deposit in the upstream cluster.
An event are selected if the event satisfy one of three following 
condition, (i) No matched EC clusters, (ii) $Edep_{\rm upstream}~>~150~{\rm MeV}$
and (iii) ${\rm R}{\rm energy}~<~0.2$.  
 
\subsubsection{Two extended track}
\label{ssxec:extrk} 
From this selection, we use the extended track information instread of the 3D track 
information. To reconstruct {\pzero}s, events with the number of extended tracks more
than one are selected. This cut is also for the dirt rejection since
there is a lot of the dirt contribution with one extended track.
In such dirt events, single gamma ray comes to SciBar and make
two 3D tracks, which is merged as one extended track.
\subsubsection{The reconstructed vertex of {\pzero}s} 
The reconstructed vertex of {\pzero}s are calculated as 
a intersection of two extended tracks. Using this information,
we select {\pzero}s produced only in SciBar to reject 
dirt events where {\pzero}s are produced at the outside of SciBar. 
Hence, the events where the reconstructed $z$-vertex of
 the \pzero is donwstream of the most upstream position of SciBar are selected.
\subsubsection{Reconstructed mass of {\pzero}s}
\label{sssec:pi0masscut} 
\begin{figure}[tbp]
   \includegraphics[keepaspectratio=false,width=50mm]{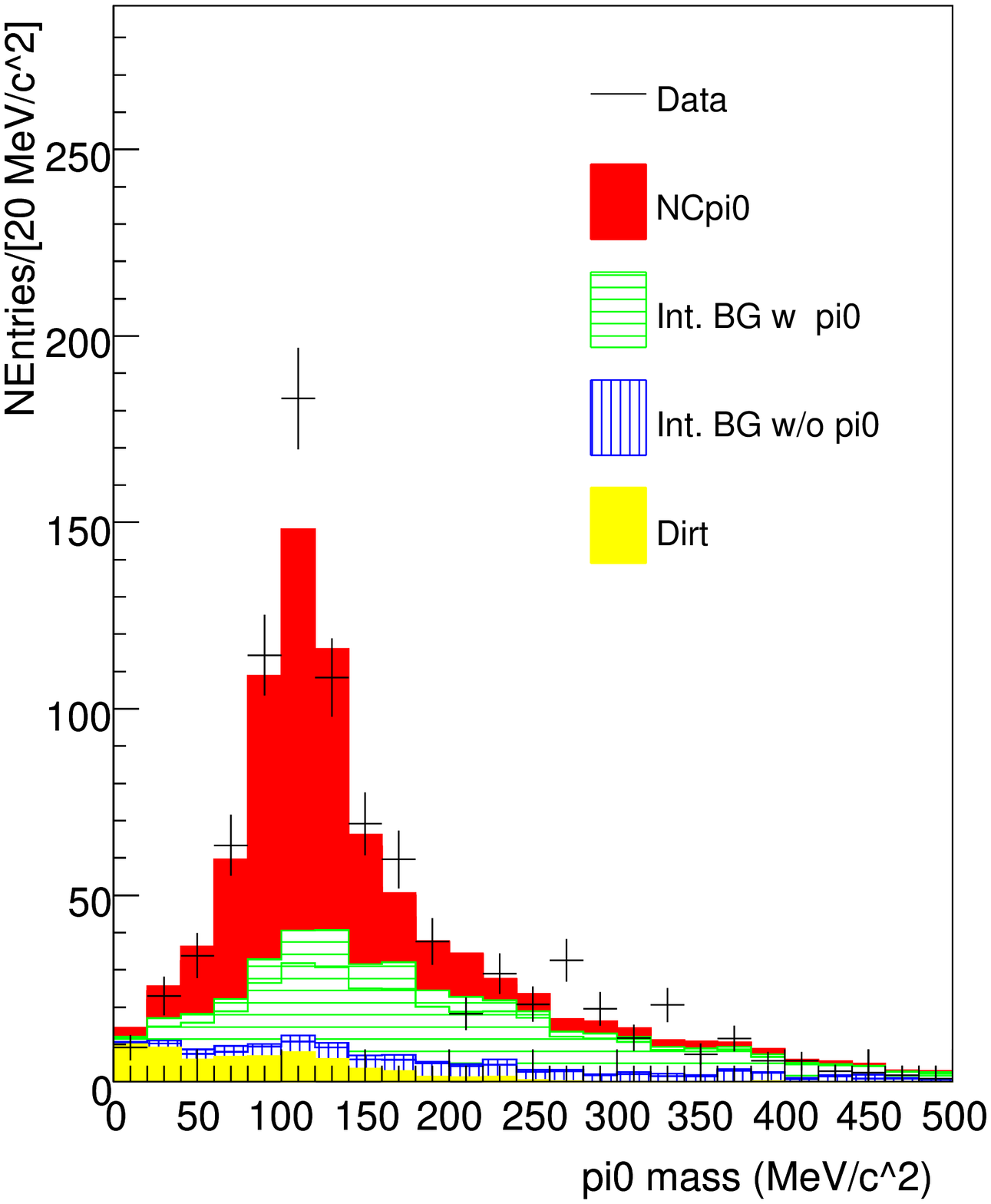}
    \includegraphics[keepaspectratio=false,width=50mm]{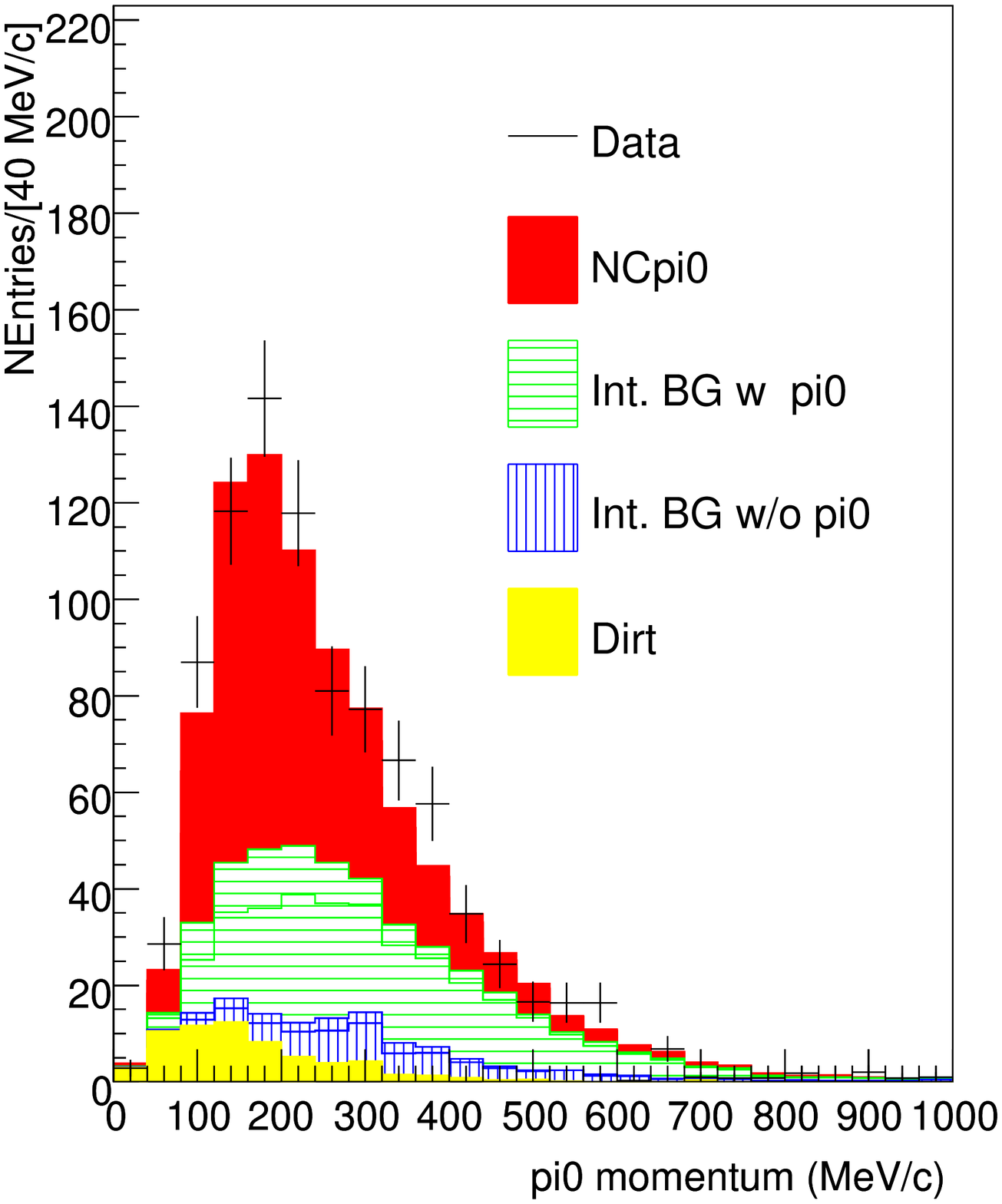}
 \caption{The reconstructed {\pzero} mass before the \pzero mass cut (left) and 
{\pzero} momentum after the \pzero mass cut (right). The contributions 
from the NC\pzero signal, the internal background with {\pzero}s in the final state, the internal background without {\pzero}s in 
the final state and the dirt events are shown separately. }
 \label{fig:pi0mass_merged}
\end{figure}

The left plot in Fig.~\ref{fig:pi0mass_merged} shows the reconstructed mass of the \pzero
calculated as   $\sqrt{2\rm{E^{rec}_{{\gamma}1}}\rm{E^{rec}_{{\gamma}2}}(1-\cos{{\theta}^{\rm{rec}}})}$
, where $\rm{E^{rec}_{{\gamma}1}}$($\rm{E^{rec}_{{\gamma}2}}$) the is energy of extended
tracks($\rm{E^{rec}_{{\gamma}1}} > \rm{E^{rec}_{{\gamma}2}}$) and
${\theta}^{\rm{rec}}$ is the 3D angle between two extended tracks. 
We select events with
 $ 50~{\rm MeV}/{c^{2}} < \rm{M^{rec}_{\pi^{\rm{0}}}} < 200~{\rm MeV}/{c^{2}} $
to reduce the background events. The fact that the peak value is smaller than
 the actual \pzero mass (135 MeV) is due to energy leakage of $\gamma$s.

We also show the reconstructed \pzero momentum after this selection
in the right plot in Fig.~\ref{fig:pi0mass_merged}.
\subsubsection{The summary of the event selections}
 \label{sssec:evsum}
Tab.~\ref{tab:evsum} shows the number of events of data and MC
 simulation at each event selection stage. We select 657 events
 after all event selections and the number of signal is estimated
 to be 374 events (after the subtraction of the secondary \pzero
 events). The purity and  efficiency of NC \pzero production after
 all event selections are estimated to be 61\% and 5.4\% , respectively.
\begin{table}[tbp]
 \caption{Event selection summary}
 \label{tab:evsum}
  \begin{tabular}{lrrrrrr}
    \hline \hline
    Event selection             & DATA   & \multicolumn{3}{c}{MC}  & NC\pzero \\
                                &        & Signal & Int. BG & Dirt BG & Efficiency \\
    \hline
    Pre-selection               & 11,926 & 1,919  & 9,782 & 895 & 27.7\%   \\
    No side escaping            &  7,444 & 1,486  & 5,686 & 638 & 21.4\% \\
    Decay-e rejection           &  5,609 & 1,396  & 3,766 & 606 & 20.1\% \\
    Trk. disconnection          &  3,614 & 1,332  & 1,688 & 595 & 19.2\% \\
    Proton rejection            &  2,123 &   745  &   943 & 408 & 10.7\% \\
    EC cut                      &  1,534 &   675  &   507 & 399 &  9.7\% \\
    Two extended trks           &    973 &   450  &   383 & 121 &  6.5\% \\
    \pzero vertex cut           &    905 &   434  &   375 &  65 &  6.2\% \\
    \pzero mass cut             &    657 &   374  &   197 &  38 &  5.4\% \\    
    \hline \hline
  \end{tabular}
\end{table}

\section{Results}
\label{sec:results}
\subsection{$\sigma(\rm{NC}\pi^{\rm{0}})/\sigma(\rm{CC})$ cross section ratio}
\label{chap:crosssection}
We measure the cross section ratio of the neutral current $\pi^{\rm{0}}$ production to
the total charged current interaction.
\subsubsection{Neutral current $\pi^{\rm{0}}$ production}
The efficiency corrected number of neutral current $\pi^{\rm{0}}$ events is calculated 
as 
\begin{eqnarray}
N({\rm{NC}}\pi^{\rm{0}}) = \frac{N_{\rm{obs}}-N_{\rm{BG}}}{{\epsilon}_{{\rm{NC}}\pi^{\rm{0}}}}
\end{eqnarray}
where $N_{\rm{obs}}$ is the number of observed events, $N_{\rm{BG}}$ is the number of
background events estimated with the MC simulation, and ${\epsilon}_{\rm{NC}\pi^{\rm{0}}}$
is the selection efficiency of neutral current $\pi^{\rm{0}}$ events calculated by the MC simulation.
$N_{\rm{obs}}$ and $N_{\rm{BG}}$,  ${\epsilon}_{\rm{NC}\pi^{\rm{0}}}$ are 657, 238.3 and 0.053, respectively. 
The mean neutrino beam energy for true neutral current  neutral pion events in the sample is estimated to be 1.16 GeV after accounting for the effects
of the selection efficiency. 
\subsubsection{Total charged current interaction}
The total number of charged current interaction is estimated by using the $MRD 
stopped$ sample. We call a 3D track in SciBar matched with a track or hits 
in the MRD as a $SciBar{\rm -}MRD~matched$ track. For the $MRD~stopped$ events, 
at least one SciBar-MRD matched track is required to stop in MRD.
The details of the selection for the MRD stopped events are described 
in \cite{Hiraide:2009} The number of  charged current candidates after 
correcting for the selection efficiency is calculated as 
\begin{eqnarray}
N({\rm CC}) = \frac{N^{CC}_{\rm{obs}} \times p_{\rm{CC}}}{{\epsilon}_{\rm CC}}
\end{eqnarray}
where $N^{\rm CC}_{\rm obs}$ is the number of observed charged current event candidates,
$\epsilon_{\rm CC}$ and $p_{\rm CC}$ are the selection efficiency and purity for charged
current interaction in the sample, respectively.
We observed 21702 MRD stopped events ($N^{CC}_{\rm{obs}}$). 
The selection efficiency and purity of charged
current events are estimated to be 19\% ($\epsilon_{\rm CC}$) and 89\% ($p_{\rm CC}$),
respectively. 
\subsubsection{Cross section ratio}
The ratio of the neutral current neutral pion production to the total
charged current cross section is measured to be
\begin{eqnarray}
\frac{\sigma({\rm{NC}}\pi^{\rm{0}})}{\sigma({\rm{CC}})}
&=& \frac{N({\rm{NC}}\pi^{\rm{0}})}{N({\rm CC})} \nonumber \\
&=& (7.7 \pm 0.5({\rm stat.})^{+0.4}_{-0.5}({\rm sys.})) \times 10^{-2}
\end{eqnarray}
at the mean neutrino energy of 1.16 GeV, where the systematic error is described later. the Neut expectation is
0.068. Therefore, the measurement is consistent with our MC simulation based on the Rein-Sehgal model 
\cite{Rein:1982pf}\cite{Rein:2006di}\cite{Rein:1980wg} for the pion production.  
\subsection{Systematic errors}
\label{subsec:systematics}
The sources of systematic error are divided into four categories,
 (i) detector response and track reconstruction, (ii) nuclear effects
 and neutrino interaction models, 
(iii) neutrino beam and (iv) dirt density. We vary these 
sources within their uncertainties and take the resulting change in the cross 
section ratio as the systematic uncertainty of the measurement. 
Tab.\ref{tab:systot} summarizes the systematic errors in the neutrial current
 neutral pion cross section ratio. The total systematic error 
is $^{+0.4}_{-0.5} \times 10^{-2}$ on the cross section ratio.
\begin{table}[htbp]
\caption{Summary of the systematic erros in the neutral current neutral pion cross section
ratio}
\begin{tabular}{ccc}
\hline
\hline
Source & \multicolumn{2}{c}{error ($\times 10^{-2}$)}\\ \hline
Detector response & -0.4 & 0.3 \\
$\nu$ interaction & -0.2 & 0.2 \\
Dirt density & -0.1 & 0.0 \\ 
$\nu$ beam & -0.1 & 0.2 \\ \hline
Total & -0.5(-0.482) & 0.4(0.411) \\
\hline
\hline
\end{tabular} 
\label{tab:systot}
\end{table}
\begin{theacknowledgments}
The SciBooNE collaboration gratefully acknowlede support from various grants, contracts 
and fellowship from the MEXT (Japan), the INFN (Italy), the Ministry of Education and Science
and CSIC (Spain), the STFC (UK). The author is grateful to the Japan Society for the Promotion
of Science for support.
\end{theacknowledgments}


\begin{thebibliography}{9}
\bibitem{Itow:2002rk}
  Y.~Itow,
  Nucl.\ Phys.\ Proc.\ Suppl.\  {\bf 112}, 3 (2002).

\bibitem{Hiraide:2009}
    K.~Hiraide {\it et al.}  [SciBooNE Collaboration] 
     Phys.\ Rev.\ D {\bf 78}, 112004 (2008).

\bibitem{Rein:1982pf}
  D.~Rein and L.~M.~Sehgal,
  Nucl.\ Phys.\  B {\bf 223}, 29 (1983).

\bibitem{Rein:2006di}
  D.~Rein and L.~M.~Sehgal,
  Phys.\ Lett.\  B {\bf 657}, 207 (2007)
  [arXiv:hep-ph/0606185].
 
\bibitem{Rein:1980wg}
 D.~Rein and L.~M.~Sehgal,
  Annals Phys.\  {\bf 133}, 79 (1981).

\bibitem{Nitta:2004nt}
  K.~Nitta {\it et al.},
  Nucl.\ Instrum.\ Meth.\  A {\bf 535} (2004) 147
  [arXiv:hep-ex/0406023].

\bibitem{Buontempo:1994yp}
  S.~Buontempo {\it et al.},
  Nucl.\ Instrum.\ Meth.\  A {\bf 349} (1994) 70.

\bibitem{Maesaka:2005aj}
  H.~Maesaka,
  Ph.D.~thesis, Kyoto University, 2005.

\end{thebibliography}
\end{document}